 \documentclass[final,1p,times]{elsarticle}

\usepackage{amssymb}
\usepackage{amsmath}

\usepackage{graphicx}
\usepackage{subfig}
\usepackage{booktabs}

\begin{document}

\begin{frontmatter}



\title{Inferring Passenger Type \\from Commuter Eigentravel Matrices}


\author{E.F. Legara and C. Monterola}

\address{Institute of High Performance Computing, \\
Agency for Science, Technology, and Research, \\Singapore 138632}

\begin{abstract}
A sufficient knowledge of the demographics of a commuting public is essential in formulating and implementing more targeted transportation policies, as commuters exhibit different ways of traveling---including time in the day of travel, the duration of travel, and even the choice of transport mode. With the advent of the Automated Fare Collection system (AFC), probing the travel patterns of commuters has become less invasive and more accessible. Consequently, numerous transport studies related to human mobility have shown that these observed patterns allow one to pair individuals with locations and/or activities at certain times of the day. However, classifying commuters using their travel signatures is yet to be thoroughly examined.  

Here, we contribute to the literature by demonstrating a procedure to characterize passenger types (Adult, Child/Student, and Senior Citizen) based on their three-month travel patterns taken from a smart fare card system. We first establish a method to construct distinct commuter matrices, which we refer to as \emph{eigentravel} matrices, that capture the characteristic travel routines of individuals. From the eigentravel matrices, we build classification models that predict the type of passengers traveling. Among the models explored, the gradient boosting method (GBM) gives the best prediction accuracy at $76\%$, which is $84\%$ better than the minimum model accuracy (41\%) required vis-\`a-vis the proportional chance criterion. In addition, we find that travel features generated during weekdays have greater predictive power than those on weekends. This work should not only be useful for transport planners, but for market researchers as well. With the awareness of which commuter types are traveling, ads, service announcements, and surveys, among others, can be made more targeted spatiotemporally. Finally, our framework should be effective in creating synthetic populations for use in real-world simulations that involve a metropolitan's public transport system. 

\end{abstract}


\end{frontmatter}


\section{Introduction}
The era of big and smart data has provided a substantial impetus in understanding human mobility---revealing the regularity and predictability of human behavior. In transport studies in particular, the widespread use of contactless smart fare card systems has spurred considerable growth in the field~\cite{Ma2013, Lee2011,Sun2012, Chakirov2011,Legara2015}. The main focus of most disquisitions on human mobility has been on identifying and/or predicting activity locations given an individual's past transportation transactions record, essentially spotting places where an individual goes to and hangs around at certain times of day---revealing ones home, work, and ``third place''~\cite{Jarv2014,Kusakabe2014,Pelletier2011,Goulias1999,Nassir2015,Lee2014}. 

Understanding human mobility is especially consequential in urban land-use and transportation planning~\cite{Chu2008,Utsunomiya2006}.  Gaining insights on where people go and what activities they engage in, or even inferring what drives them to travel from one place to another, can help in designing smart cities that can sufficiently address the needs of their citizens from their environment~\cite{Medina2014, Othman2014}; thereby improving their overall well-being.

Notwithstanding the fact that most human mobility studies are centered on matching individuals with locations and/or activities, certain sociotechnical datasets have more to offer other than spatial information. In this study, for example, we utilize data from travel fare cards that not only have spatiotemporal information such as origin, destination, time of travel, and duration of travel, but also provide a particular demographic information, which is the type of passenger traveling, i.e. Adult, Child/Student, or Senior Citizen. In this work, instead of predicting where people go at certain times of the day, we determine a set of features based on travel routines that can help identify \emph{which} passenger types are traveling. Realizing commuter types can give us a better understanding of the structure of a society and the needs of its people from their surroundings. From the perspective of transport planning, this can help stakeholders quantify more systematically how a certain group of commuters would react to or be affected by changes in the entire transport system---from infrastructure changes to policy changes~\cite{Medina2014,Tong2013}. Finally, from the standpoint of modeling and simulations, our proposed approach can aid in setting up synthetic populations wherein different passenger categories exhibit varying travel signatures. 

The paper is organized as follows. In the next section, we discuss the data used in the study. This is then followed by a methods section where (1) present some descriptive statistics relevant to the construction of our classification models, and (2) demonstrate in detail how we set up the eigentravel matrices that define the feature variables used in building the classification models. Finally, we end the article with a discussion and conclusion section where we elaborate on our results and share some insights into them.

\section{Data}

This  paper  looks into movements of public transport commuters within Singapore using a three-month travel dataset. In the city-state, there is only one smart fare card system called EZ-link used in both its bus and rail transit system (RTS). Moreover, the public transport system has both entry and exit automated fare collection (AFC) for the bus and RTS. With both entry and exit AFC, the durations of travel for each transaction can be evaluated in a straightforward manner. 

The dataset at hand has more than 3 million unique and anonymized card ID's; this includes single journey transactions across the three months under study. For purpose of computation, we  utilized  a randomly sampled population of 30,000 \emph{regular} commuters. Such sampling yields a confidence interval equal to 99.99\% or an error of less than 0.01\%. The population is equally split among three passenger types: adult, child/student, senior citizen. 


Each travel transaction contains the following pieces of information that are relevant to the study: card ID, origin, destination, start date (of travel), start time (of travel), end time (of travel), mode of transport (bus or rail), and passenger type.

\section{Methodology}
\label{sec:methods}

\subsection{Descriptive Statistics}
\label{sec:distanal}

\begin{figure}[ht]
\centering
\begin{tabular}{cc}
\multicolumn{2}{c}{\includegraphics[width=100mm]{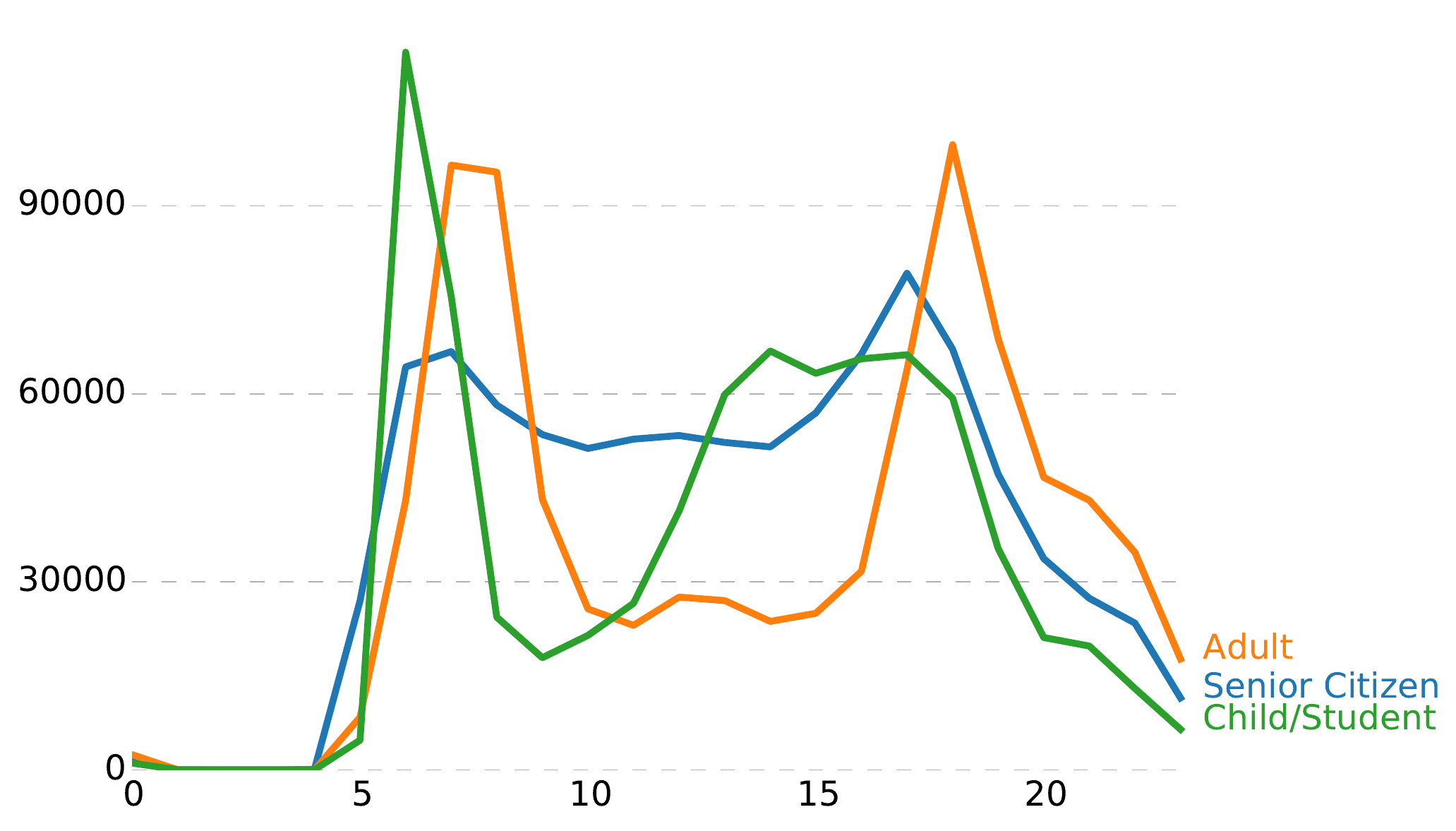} }\\
\multicolumn{2}{c}{(a) Weekdays}\\[6pt]
  \includegraphics[width=65mm]{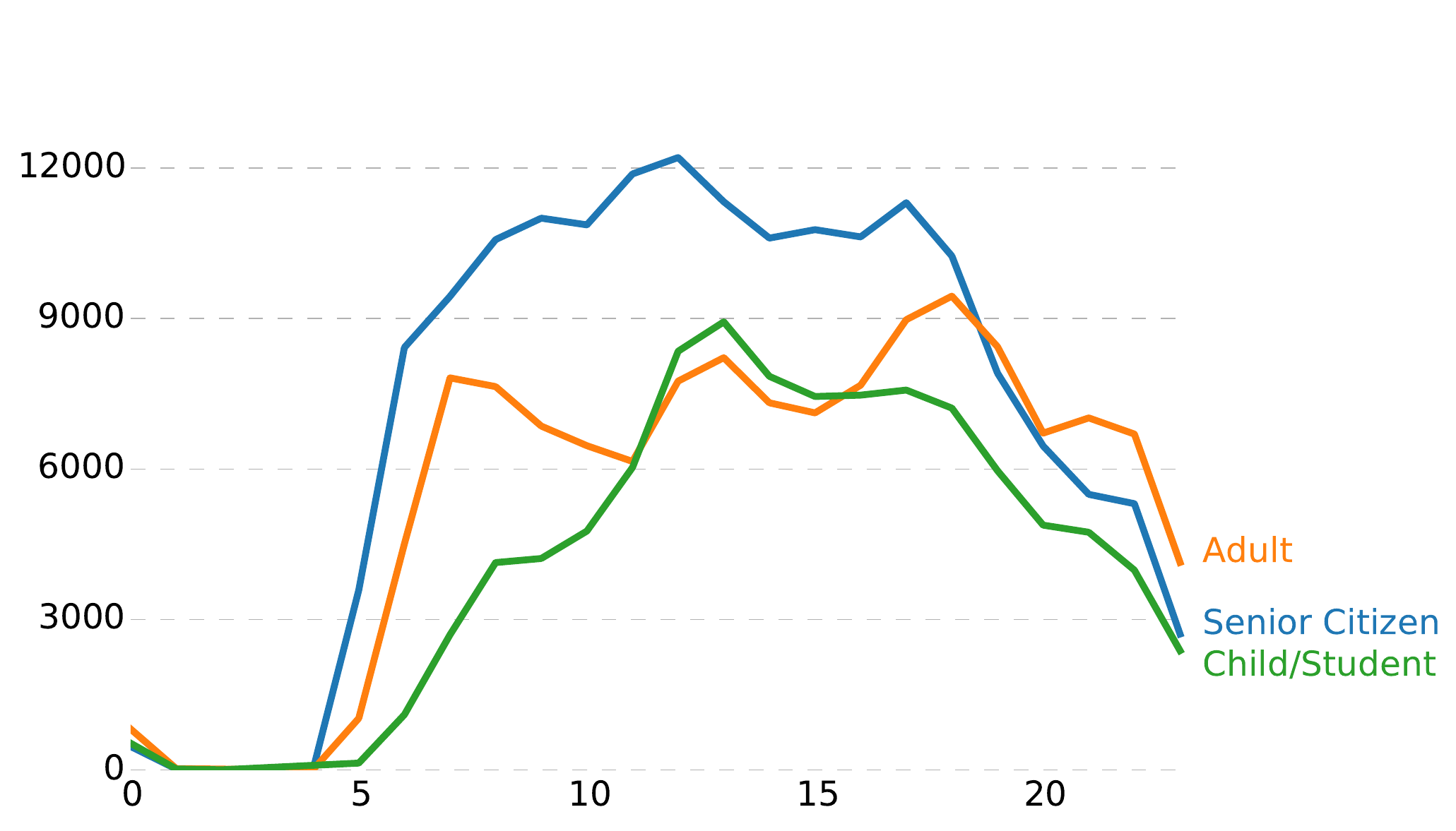} &   \includegraphics[width=65mm]{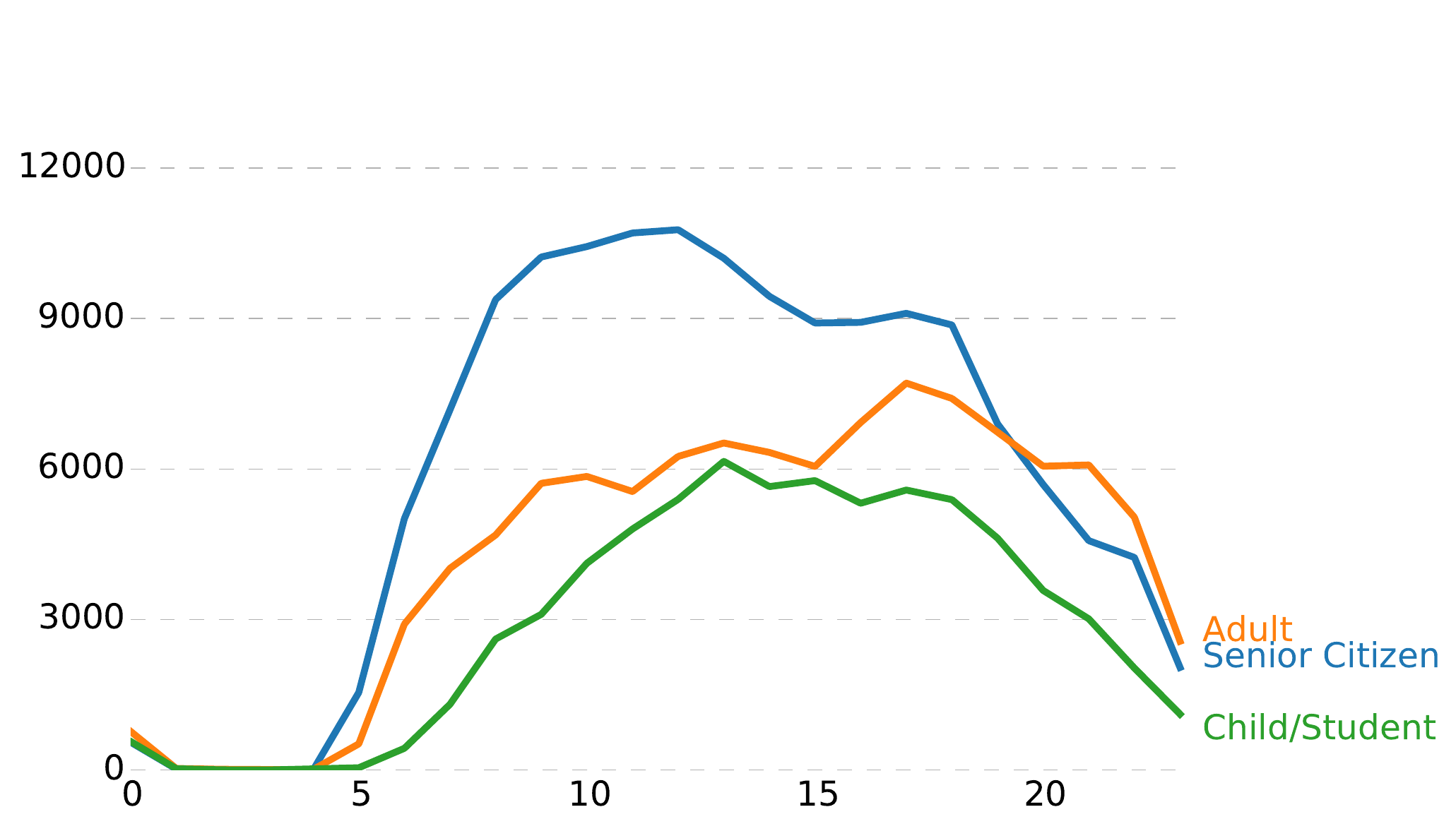} \\
(b) Saturdays & (c) Sundays \\
\end{tabular}
  \caption{\textbf{Travel Demand Curve}. Three different travel demand curves are plotted. One for the ``Adult'' population, another for the ``Child/Student'', and finally for the ``Senior Citizen'' population. The ``Adult'' demand curve displays the typical demand curves that are discussed in the transportation research literature wherein there are two distinct peaks---one for the morning peak hours and another for the evening peak hours. On the other hand, the ``Child/Student'' demand curve only exhibits one well-defined peak. Finally, the ``Senior Citizen'' curve displays no distinquishable peak, but instead a plataeu suggesting that senior citizens typically do not follow a ``universal'' routine wherein they go to work in the mornings and go home from work in the evenings.}
\label{fig:ridestarttime}
\end{figure}

\begin{figure}[ht]
\centering
\begin{tabular}{cc}
\multicolumn{2}{c}{\includegraphics[width=65mm]{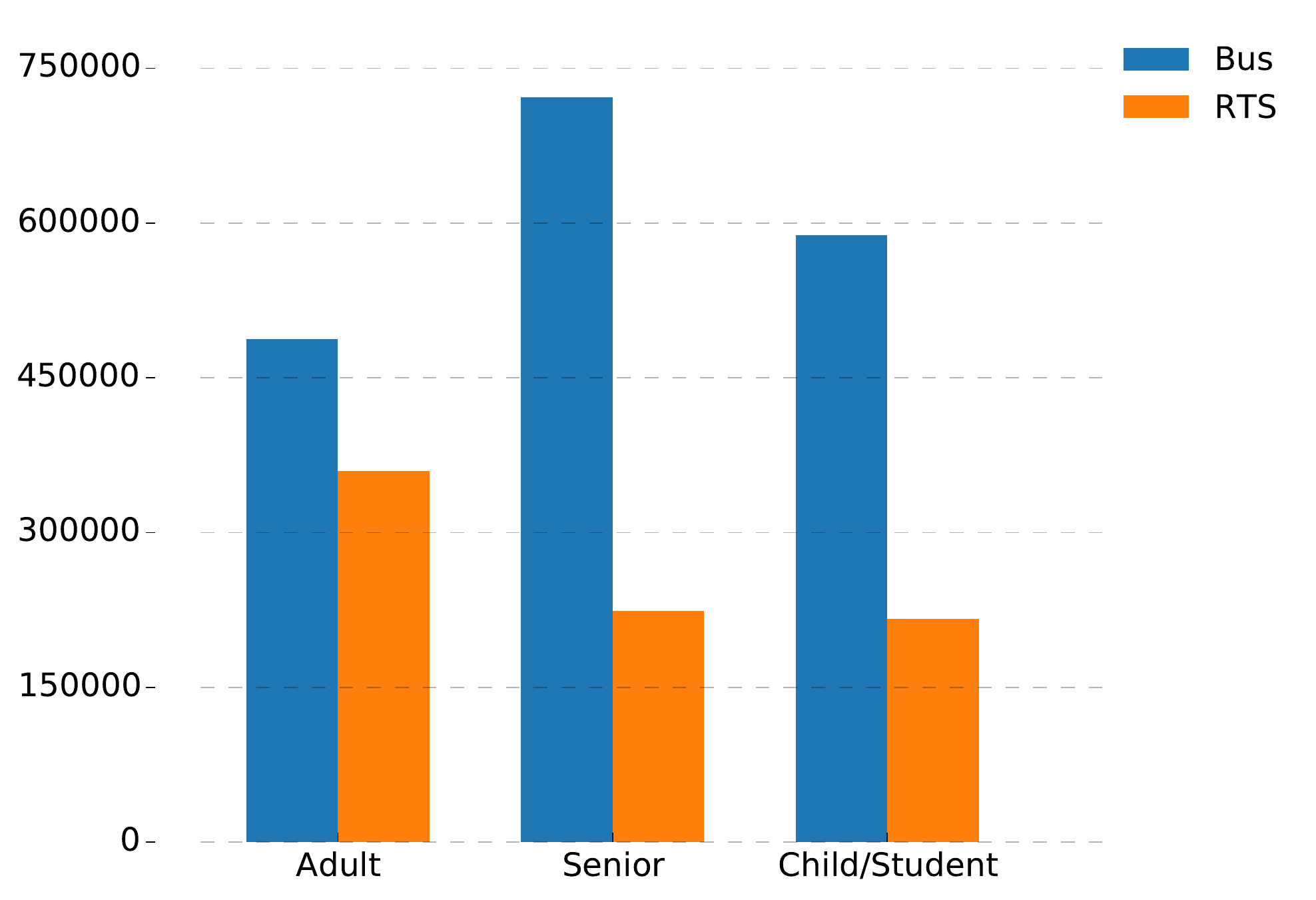} }\\
\multicolumn{2}{c}{(a) Weekdays}\\[4pt]
  \includegraphics[width=65mm]{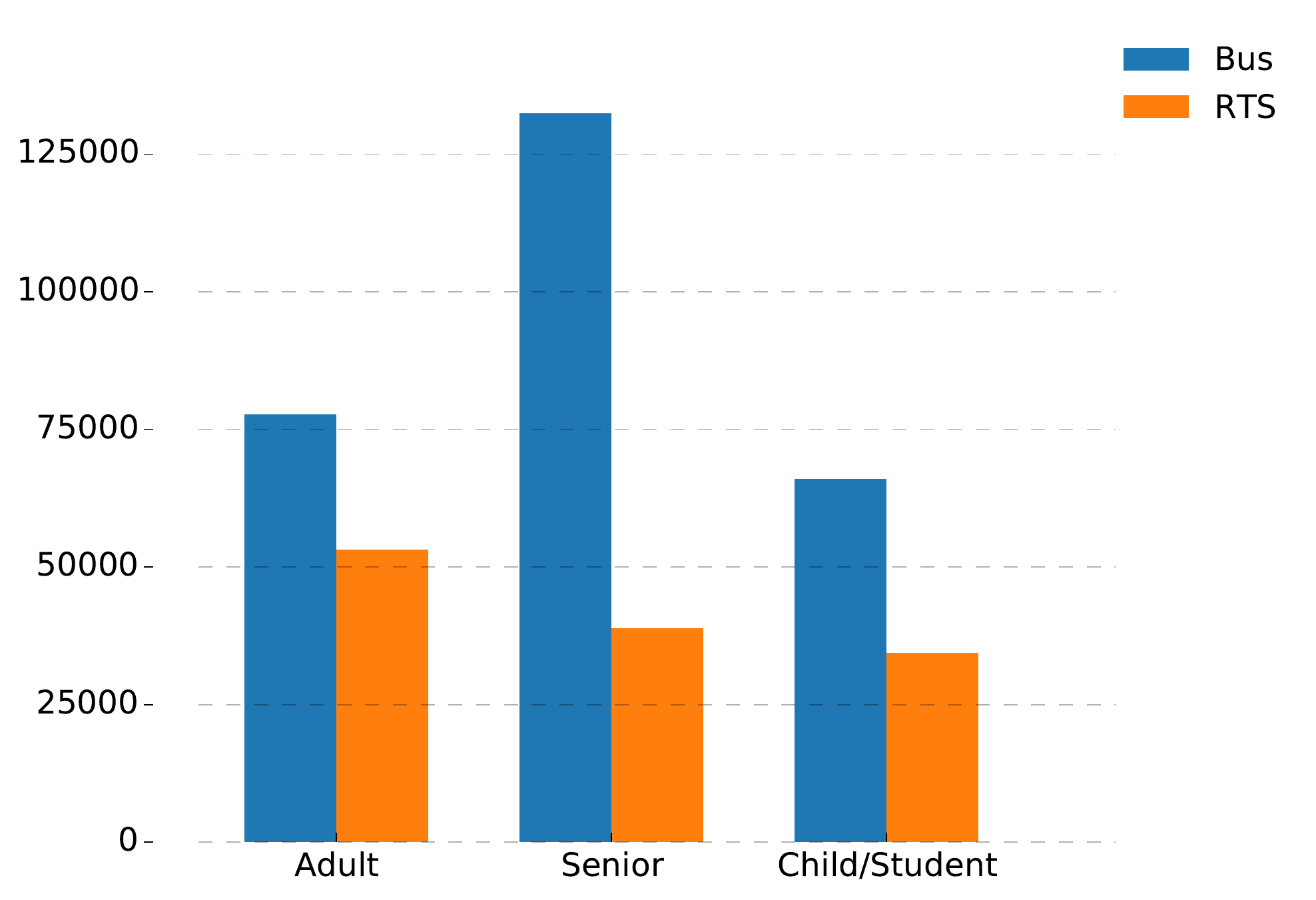} &   \includegraphics[width=65mm]{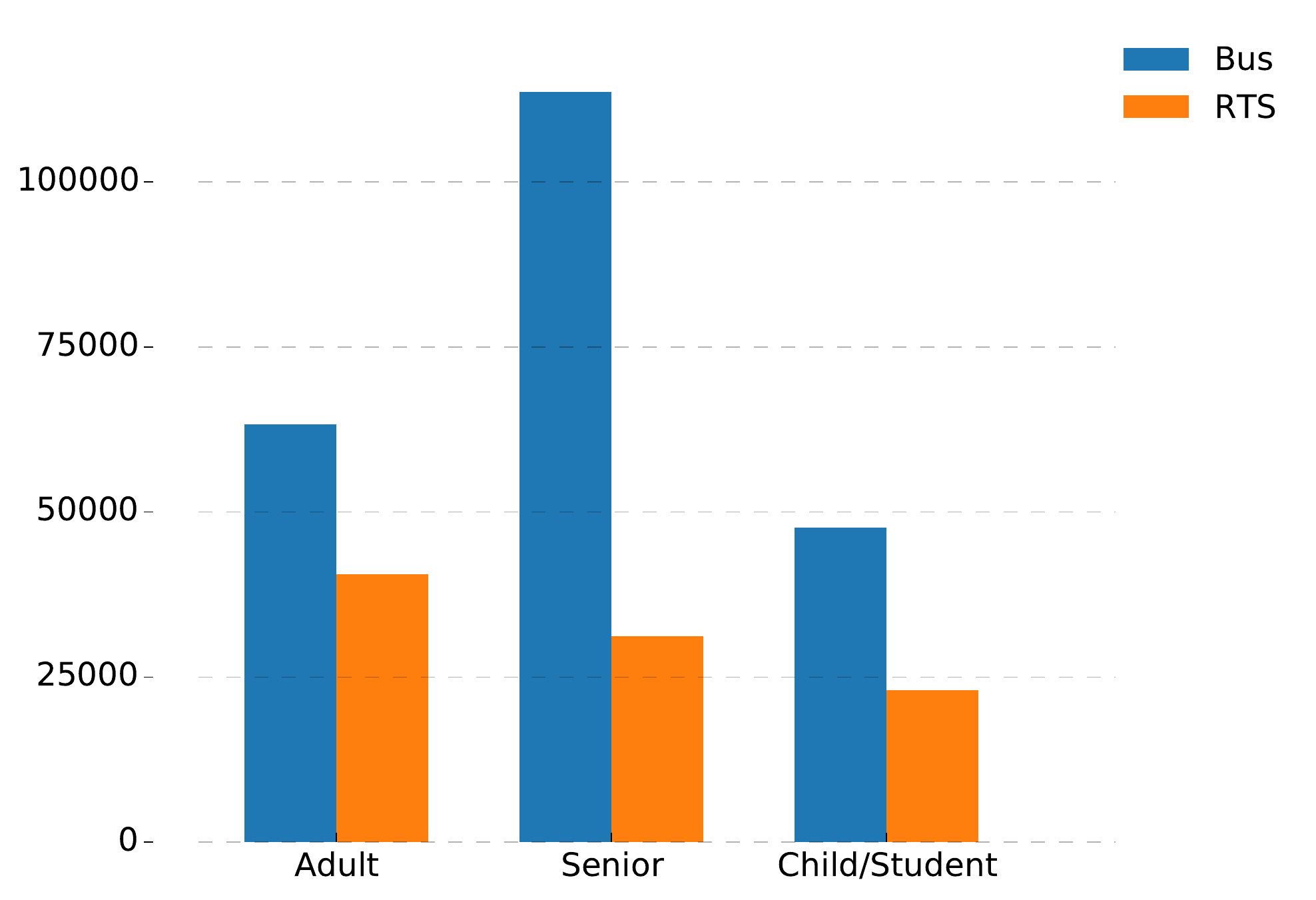} \\
(b) Saturdays & (c) Sundays \\
\end{tabular}
  \caption{\textbf{Mode of Transport}. For the three subplots (a), (b), and (c), the travel mode distributions for three types of passenger population are shown. Except for the drop in the ridership on weekends, the mode trends are quite consistent across the week, i.e. the bus ridership dominates the RTS. The dominance is especially evident for the ``Senior'' population were more than $50\%$ over the total trips utilize the bus system. The ``Adult'' population utilizes the RTS more often than the other two.}
\label{fig:transportmode}
\end{figure}

We first look at some descriptive statistics that can be derived from the dataset. In Fig.~\ref{fig:ridestarttime}, we show the temporal travel demand statistics via the ride start time distributions for each passenger type. A typical weekday travel demand curve that we see in the literature is that there are two distinct peaks that correspond to both the AM and PM peak hours--- when people go to work/school (in the morning) and when they go home (in the afternoon)~\cite{Sun2012, Legara2015}. However, when we discriminate across passenger types, we see three distinct curves (Fig.~\ref{fig:ridestarttime}a). The curve for the passenger type ``Adult'' (\emph{A}-curve) is the same as the usual travel demand curves presented in the literature. However, for the travel demand curve of children/students (\emph{C}-curve), we see that there is only one sharp peak that is found in the morning; in the afternoon, the curve plateaus. This suggests that children/students have practically varying end-of-school times---spread from 1300 hours to 1800 hours as there are students who only go to class in the morning. Finally, the travel demand curve for the elderlies (senior citizen, \emph{S}-curve) does not reveal a peak, which implies that seniors typically do not have a ``universal'' schedule. These three demand curves give a hint on how to set up the different travel features for classifying passenger types. We probe these curves in greater detail in the Results and Discussion section below.

We also look at how the two modes of Singapore public transport, bus and train, are utilized across the period under study for the three types (\emph{see} Figure~\ref{fig:transportmode}). The barplots show that, in general, the usage of bus dominates that of the rail. This is more pronounced for the elderlies wherein around $x\%$ of the total trips account for the bus usage. 

\subsection{Eigentravel Matrices} \label{sec:bematrices}

\begin{figure}[h!]
  \caption{\textbf{Eigentravel Matrix $B$: Schematic Diagram}. The eigentravel matrix $B$ is a $42 \times 20$ matrix. The entire dataset covers a total of fourteen (14) weeks. We then separate the weekdays from the weekends, and further distinguish between saturdays and sundays. The first fourteen rows are for the fourteen weeks for the weekdays, while the last two fourteen-week slabs, for saturdays and sundays. Each row is referred to as a $w$-slice or a week-slice. The twenty columns, on the other hand, represent each hour from 0400 hours to 2359 hours of each day. Each cell or hour slice in a $w$-slice is referred to as one $h$-slice. For the first fourteen rows, the characteristic travel patterns are averaged across the weekdays of each week. Finally, each $b_{w,h}$ cell has a value that ranges from 0 to 10. Details are discussed in Fig.~\ref{fig:behaviormatrixweek}.}
  \centering
    \includegraphics[width=3.5in]{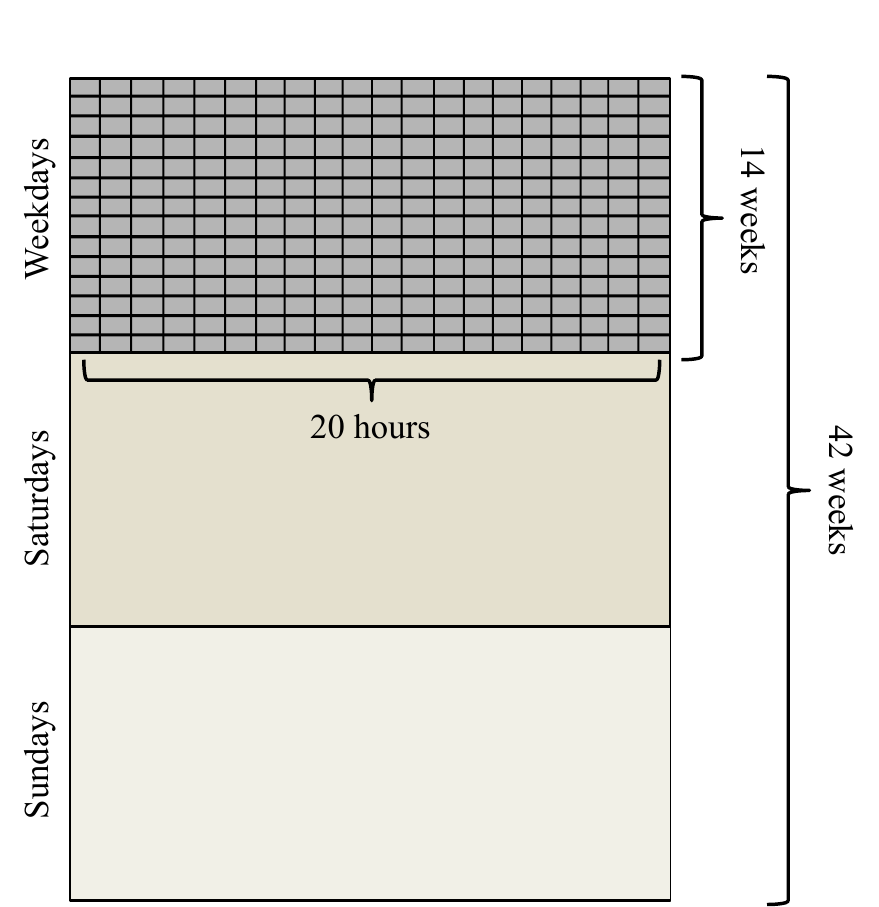}
	\label{fig:behaviormatrix}
\end{figure}

Building from what have been established in the previous section, we construct a unique eigentravel matrix $B_i$ for each agent $i$ to characterize an individual's travelprint. $B_i$ captures, at the minimum, the observed differences in travel demand (or ride times) of each passenger type and their preferred modes of transport. 

\begin{figure}[h!]
  \caption{\textbf{$w$-slice: Eigentravel Pattern for a Week Slice.} Each row in matrix $B$ are divided into 20 $h$-slices corresponding to each hour from 0400 hours to 2359 hours. Each $h$-slice is further divided into 60 $m$-slices or minute slices. In this figure, we illustrate how a set of travel transactions of an individual as shown in Table~\ref{tab:transactions} is represented in $B$. Journey $1$ involves a total of $\Delta \rho =12$ minutes of travel by bus from 0651 to 0702 hours. For Journey $1$, $h$-slices 3 and 4 are partially shaded accrodingly (with blue for the bus travel mode)---a total of 9 $m$-slices for $h=3$, and 2 $m$-slices for $h=4$. Similarly, for Journey $2$, $h$-slices 15 and 16 are partially shaded (red for the use of the RTS) covering 18 $m$-slices and 17 $m$-slices, respectively. Finally, for Journey 3, $\Delta \rho$ = 6 $m$-slices are shaded in $h=16$ for the bus trip from 1953 hours to 1959 hours. Note that since the focus of the study is from 0400 hours, $h=1$ correspond to 4:00 AM, $h=2$ to 5:00 AM, and so on.}
  \centering
    \includegraphics[width=5.5in]{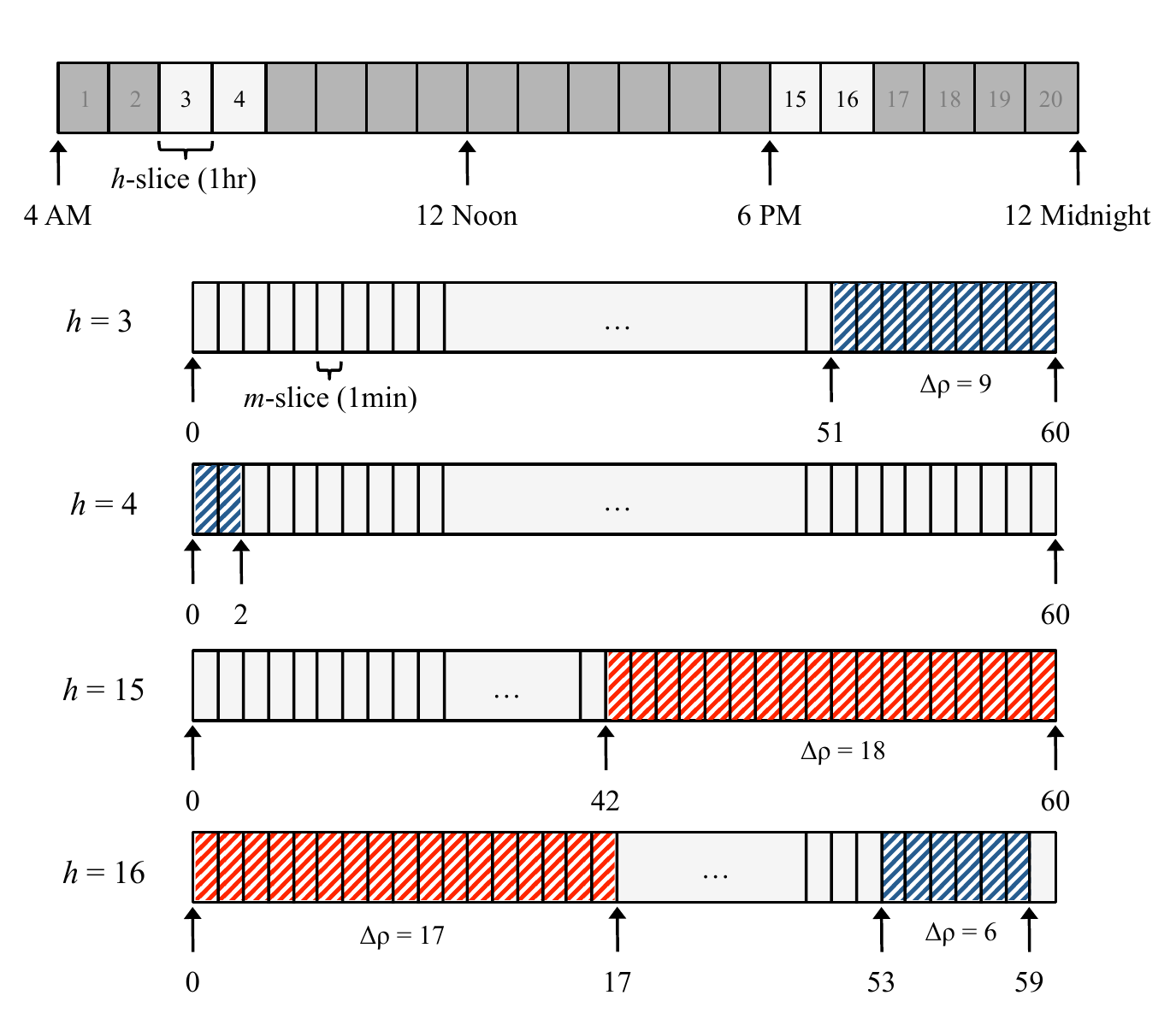}
	\label{fig:behaviormatrixweek}
\end{figure}

$B_i$ is a two-dimensional $42 \times 20$ matrix. The forty-two (42) rows correspond to three 14-week partitions from the three-month data. The first fourteen rows aim to capture the travel patterns on weekdays; while the second and third fourteen-week slabs correspond to saturdays and sundays, respectively. In this study, only trips between 0400 and 2359 hours of each day are captured--- this is depicted in the 20 columns that represent each hour of the time period under study. Figure~\ref{fig:behaviormatrix} shows a schematic diagram of an individual $i$'s $B_i$-matrix. In Figure~\ref{fig:behaviormatrixweek}, we zoom in on one of the forty-two week slices in $B_i$. The figure is discussed in greater detail below. 

Meanwhile, the eigentravel matrix $B_i$ is constructed to not only quantify \emph{when} an agent is traveling, but to also carry information on a commuter's transport mode of choice and his/her durations of travel. Each cell $b_{w,h}$ in $B_i$ can have a value in the range $[0,10]$ and is given by

\begin{equation}
  b_{w,h}=  \frac{\sum _{j \in J_w^h} f \Delta \rho }{ 60} 
\begin{cases}
f = 1, \texttt{   bus} \\
f= 10, \texttt{  train}
  \end{cases}
\end{equation}

\noindent
where $j = 1, 2, ...$ is a journey in the journey set $J_w^h$, which is a collection of all journeys that begin on week $w$ at hour $h_0$ and ends on the same week and day at hour $h_f$ where $h_f \geq h_0$.  $\Delta \rho$, on the other hand, is the duration of travel (in minutes) of the individual in week $w$ and hour $h$. If a journey covers two adjacent hours, say the travel was from 0651 hours to 0702 hours covering hours 0600 ($h=3$) and 0700 ($h=4$), respectively, the corresponding travel duration for each hour will be counted separately. Finally, to distinguish between using a bus and a rail transit, a multiplier $f$ of either 1.0 or 10.0 is introduced.

\begin{table}[h]
\caption{\textbf{Sample travel transactions of a commuter.} The table shows three hypothetical journeys taken by a commuter. We use this to illustrate how we build an individual's \emph{eigentravel} matrix. The table entries are ``visualized' in Figure~\ref{fig:behaviormatrixweek}}.
\centering
\begin{tabular}{llllll}
\hline
{\bf Journey} & {\bf \begin{tabular}[c]{@{}l@{}}Start \\ Date\end{tabular}} & {\bf \begin{tabular}[c]{@{}l@{}}Start \\ Time\end{tabular}} & {\bf \begin{tabular}[c]{@{}l@{}}End \\ Time\end{tabular}} & {\bf \begin{tabular}[c]{@{}l@{}}Travel\\ Mode\end{tabular}} & {\bf \begin{tabular}[c]{@{}l@{}}Passenger\\ Type\end{tabular}} \\ \hline
1       & Monday, Week 1                                              & 6:51 AM                                                     & 7:02 AM                                                   & Bus                                                         & Adult                \\
2       & Monday, Week 1                                              & 6:42 PM                                                     & 7:17 PM                                                   & RTS                                                         & Adult                \\ 
3       & Monday, Week 1                                              & 7:53 PM                                                     & 7:59 PM                                                   & Bus                                                         & Adult                \\ \hline
\end{tabular}
\label{tab:transactions}
\end{table}

\begin{figure}[h!]
  \caption{{\bf Eigentravel Matrices.} Randomly sampled eigentravel matrices for illustration. Three samples for each passenger type (one per row).}
  \centering
    \includegraphics[width=5.5in]{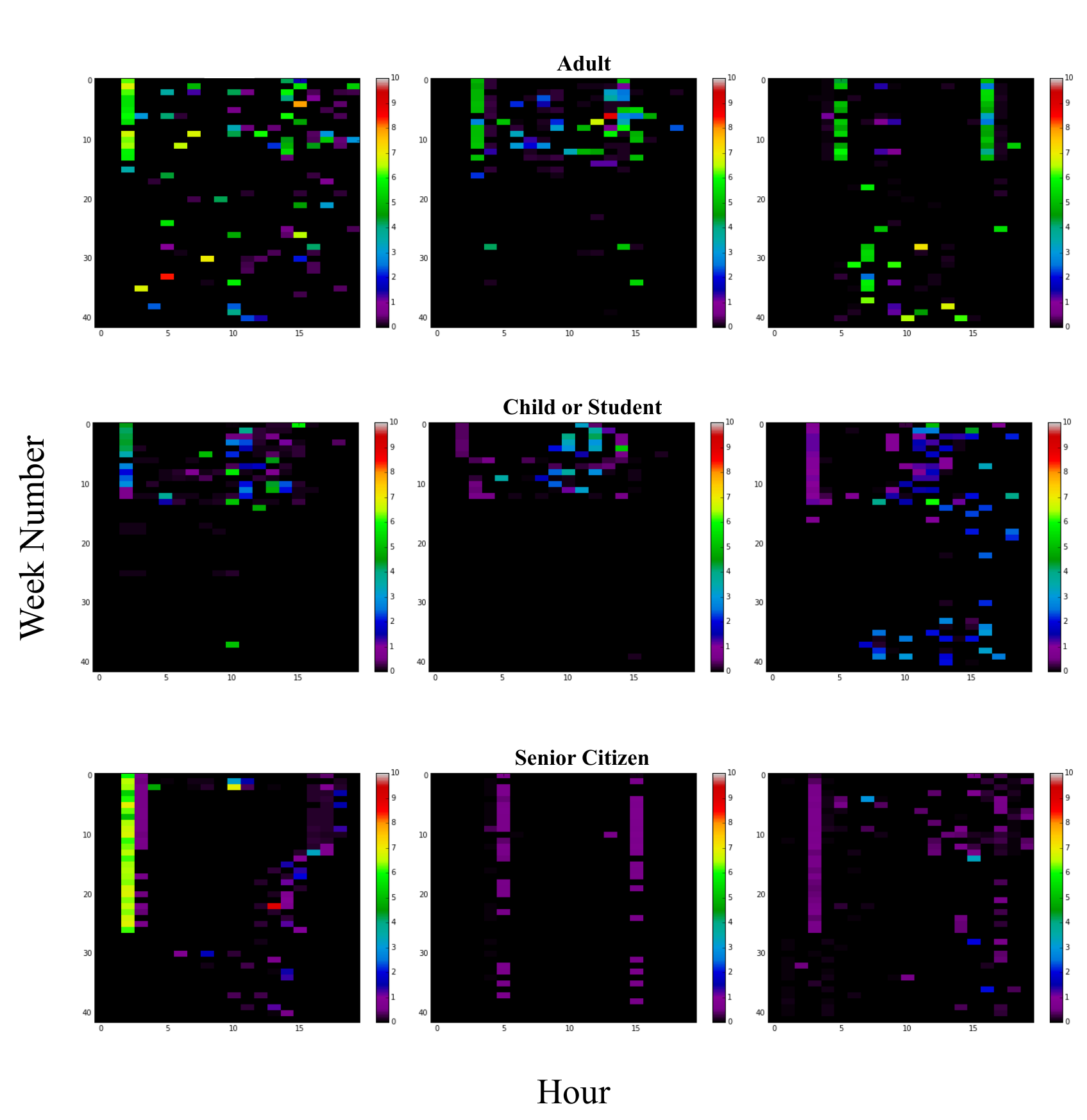}
\label{fig:samples}
\end{figure}

To illustrate the construction of $B_i$, consider the travel transactions of a hypothetical agent in Table~\ref{tab:transactions}. In Journeys 1 and 3, the agent utilized the bus system; therefore, $f=1$ for the two trips. For Journey $2$, on the other hand, the factor $f=10$ since the agent utilized the rail transit system (RTS). For Journey $1$, the trip crosses two \emph{h}-slices---$h=3$ and 4, respectively (see Fig.~\ref{fig:behaviormatrixweek}). Note that in this study, we are starting at 0400 hours ($h=1$), therefore, $h=3$ and $h=4$ for 6AM and 7AM, respectively. In Fig.~\ref{fig:behaviormatrixweek}, we can see that Journey 1 covers a duration of $\Delta \rho=9$ mins and $\Delta \rho = 2$ mins for $h=3$ and $h=4$, respectively. Consequently, cells $b_{1,3}$ and $b_{1,4}$ of $B_i$ will have non-zero values, and are computed as follows:  $b_{1,3} = \frac{1 \times 9}{60}$ and $b_{1,4} = \frac{1 \times 2}{60}$. From Journey 2, $b_{1,15} = \frac{10 \times 18}{60}$. Finally, from Journeys 2 and 3, $b_{1,16} = \frac{(10 \times 17) + (1 \times 6)}{60}$. Actual samples of eigentravel matrices are shown in Figure~\ref{fig:samples}.

\subsection{Classification}

We utilize  different supervised machine learning models and perform predictive analytics on the constructed eigentravel matrices. The three best models are (1) a distributed random forest (DRF) model, (2) a gradient boosting method (GBM), and (3) a support vector machine (SVM). These methods are standard advanced classification techniques in machine learning and have demonstrated success in a wide range of systems~\cite{Etter2012,Zhang2015,Hastie2009,Leshem2007}. 

Both DRF and GBM are forward-learning ensemble models made-up of multiple basis elements---the decision trees (DT)~\cite{Zhang2015,Friedman2001,Click2015}. Each DT in each of the ensemble provides a ``weak'' solution to the classification problem at hand. The main difference between DRF and GBM lies in how the two models generate their base models. In the DRF, the individual DTs are generated indepedently, and the fitting simply averages the performance of each of the learners; in the GBM, on the other hand, a gradient-descent based boosting formulation  with the objective of minimizing the loss function in every iteration is implemented in spawning new learners. In spite of this, similar to DRF, the final fitting is just the average of the base models. Finally, SVM is a supervised classification technique where sample clusters are separated by defining hyperplanes that give the largest minimum distances from each cluster.

The forward-learning ensemble models DRF and GBM are implemented using the H2O Python Module~\cite{h2o}, while the SVM is performed using 
$\texttt{scikit-learn}$~\cite{scikit}---a machine learning Python module. We note that linear models and deep learning methods produce results that are inferior when compared to the methods described above.

\subsection{Features}

We reshape each of the eigentravel matrices into one-dimendional arrays whose elements correspond to the $42 \times 20 = 840$ features considered in this study. The features contain information on the travel time of the individuals and their preferred mode of transport as described in Section~\ref{sec:bematrices}. The predictor variables are labelled $F1, F2, ..., F840$, where $F1$ corresponds to the average ride pattern of an individual during weekdays for the first week under consideration at 0400 hours. $F2$, on the other hand, is for the same week averaged across weekdays at 0500 hours. Finally, $F400$ is the 5th Saturday in the data set at 2400 hours. The response variable is the \emph{passenger type}.

\section{Results and Discussion}

We compare our results for each of the models by comparing their accuracy rates against the proportional chance criterion (PCC)---a common yardstick in evaluating the success of a classifier when compared to a random chance prediction~\cite{Legara2011}. PCC is calculated by summing the squared proportion of each of the group represented in the sample. As a rule of thumb, a successful model, indicative of a significant predictive score, should have an accuracy of at least $25\%$ of the PCC~\cite{Hair2009}. Accordingly, our objective is to have an accuracy of at least $PCC \times 1.25 =  [(  \frac{75}{225}  ) ^2 *3  ] \times 1.25 = 0.4125$. 

\begin{figure}[h!]
  \caption{{\bf Variable Importance Heat Maps}.  The 1 (bottom) to 42 (top) rows represent the week number (1-14 for weekdays, 15-28 for saturdays, and 29-42 for sundays) while the columns represent the 20 hours under consideration. The boxed portions of the maps highlight the variables associated with weekday travels.}
  \centering
    \includegraphics[width=4.5in]{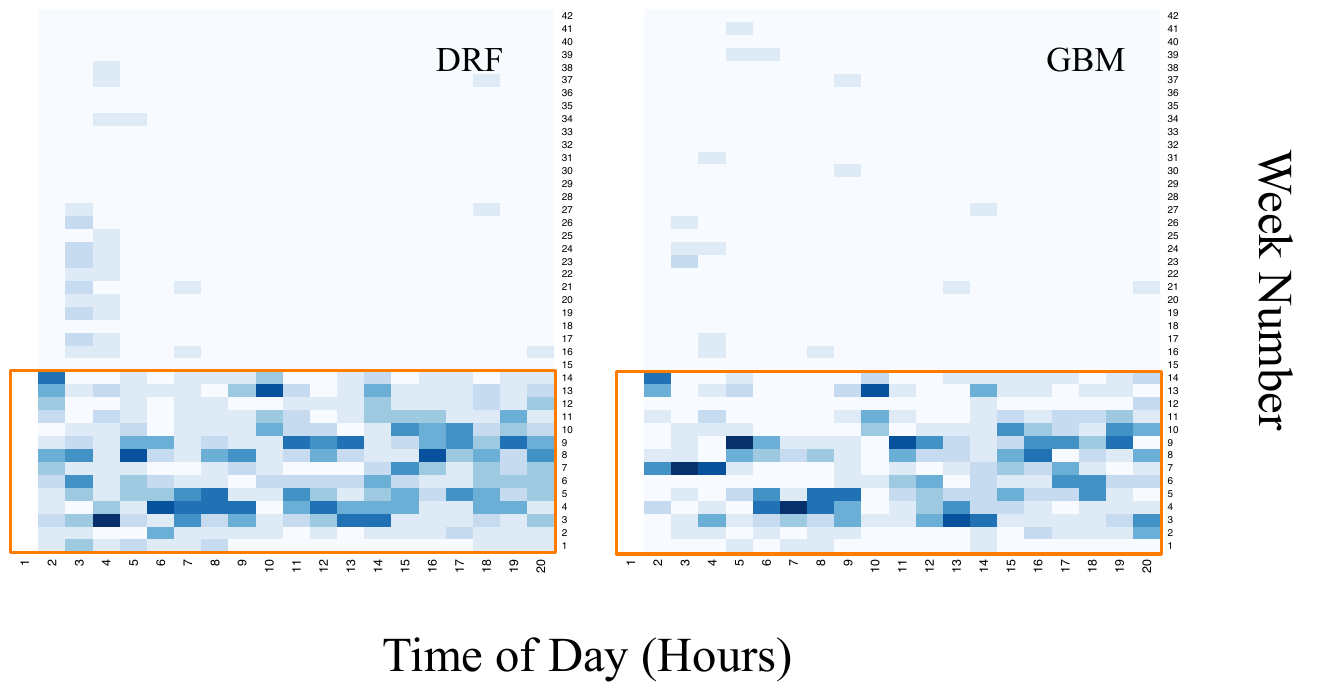}
\label{fig:varimpt}
\end{figure}

Among the three models, GBM resulted to the highest prediction accuracy of 76\%, which is 84\% better than the minimum required model accuracy (41\%) derived from the PCC. DRF and SVM gave 72\% and 64\% accuracy rates, respectively. The deep learning method we performed with layers sampled from 1 to 200 and hidden nodes from 100 to 600 only reached a maximum of 64\%, while results from the linear methods are just within $1.25 \times PCC$.

\begin{figure}[h!]
\caption{\textbf{Mean Scaled Variable Importance Across Weekdays}. The plot shows the average scaled variable importance for each hour across weekdays. Two methods are highlighted herewith: distributed random forest (DRF) and gradient boosting machine (GBM). What can be seen here is that for the GBM, the hours 0600, 1100, 1400, and 1500 dominate the other predictor variables; for the DRF, on the other hand, hours 0800, 1100, 1400, and 1500 govern the others. }
  \centering
    \includegraphics[width=3in]{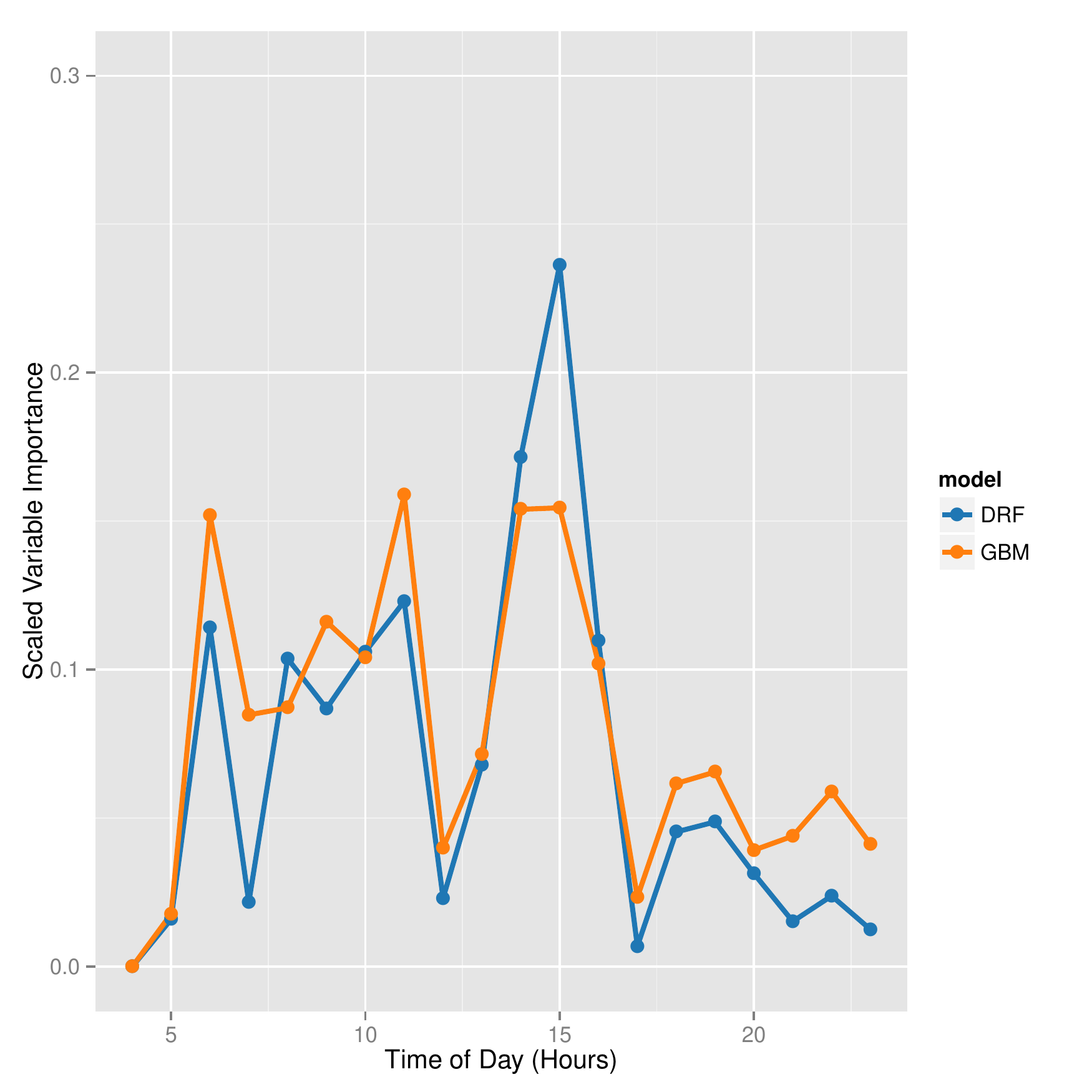}
\label{fig:meanvarimpt}
\end{figure}

Focusing on both GBM and DRF, which resulted to greater than 70\% accuracy rates, we provide heatmaps of the scaled variable importances (see Figure~\ref{fig:varimpt}). What is apparent in the figure is that most of the variables associated with the trips made during weekdays (boxed slabs) dominate the rest of the features; that is, the predictor variables corresponding to weekend travels do not contribute significantly in identifying passenger clusters. This finding concurs with the travel demand curves shown in Fig.~\ref{fig:ridestarttime} where the overall profiles of the curves for the three types for both Saturdays (Fig.~\ref{fig:ridestarttime}b) and Sundays (Fig.~\ref{fig:ridestarttime}c) are structurally similar.

In Fig.~\ref{fig:meanvarimpt}, we zoom in on the weekdays of the fourteen weeks by taking the average of the scaled variable importances of features that represent the same hour of the weekdays. In the plot, for the GBM, the leading variables are those identified with the hours: 0600, 1100, 1400, and 1500; for the DRF, the prominent variables are those at 0800, 1100, 1400, and 1500. 

\begin{table}[h!]
\centering
\caption{{\bf Confusion Matrix}. This is the confusion matrix generated by the GBM. In the matrix, it is apparent that predicting children/students gives the highest accuracy,with only an error rate of approximately 15.4\%, compared to when predicting the adults and/or senior citizens. The model utilized shows difficulty in distinguishing between an adult and a senior citizen.}
\label{tab-confmat}
\begin{tabular}{llllll}
                       & {\bf Adult} & {\bf Child } & {\bf Senior } & {\bf Error}  & {\bf Rate}      \\ \hline
{\bf Adult}            & 3637         & 392                     & 971                  & 27.3\%       & 1,363 / 5,000       \\
{\bf Child} & 497         & 4223                    & 280                   & 15.4\%       & 777 / 5,000       \\
{\bf Senior}   & 1175         & 379                    & 3444                  & 31.09\%       & 1,554 / 4,998         \\ \hline
{\bf Total}            & {\bf 5,309} & {\bf 4,994}            & {\bf 4,695 }            & {\bf 24.6\%} & {\bf 3,694 / 14,998}	 \\ \hline
\end{tabular}
\end{table}

The variable importance values may be explained by looking at Fig.~\ref{fig:ridestarttime} and Table~\ref{tab-confmat}, which shows a sample confusion matrix resulting from implementing the GBM. Table~\ref{tab-confmat} establishes that predicting children and/or students gives the highest accuracy with only an error rate of approximately 17.4\%; compared to the adults and senior citizens where the error rates are 28.5\% and 34.4\%, respectively. In addition, most of the misclassification are between the adults and senior citizens; therefore, we reckon that predictor variables that maximize the dissimilarity between the adults and senior citizens will play more significant roles in the models.

We now discuss what insights we can derive from the travel patterns of the commuters, focusing on the sets of the most relevant predictor variables. For ease of discussion, we introduce $v_h$ to represent a set of 14 weekday predictor variables that fall under a given hour ($h=1, 2, ..., 20$). To recap, from Section~\ref{sec:bematrices}, $h=1$ refers to 0400 hours, $h=2$ to 0500 hours, and $h=19$ to 2200 hours. To illustrate further, $v_{h=1} = \{F1, F21, F41, ..., F261\}$, which is a set of 14 (weekday) variables that fall within the first hour of each weekday considered. From the variable importance results for the GBM and DRF, we focus on the sets $\mathbb{G} = \{v_3, v_8, v_{11}, v_{12}\}$  and $\mathbb{D} = \{v_5, v_8, v_{11}, v_{12}\}$, respectively; these sets refer to variables under the following time frames:  0600, 1100, 1400, and 1500 hours for the GBM and 0800, 1100, 1400, and 1500 hours for the DRF. 

In Section~\ref{sec:distanal}, the general profiles of the different travel demand curves in Fig.~\ref{fig:ridestarttime} for the adults (\emph{A}-curve), the children/students (\emph{C}-curve), and the senior citizens (\emph{S}-curve) are discussed. We look into specific segments of the curves guided by the sets $\mathbb{G}$ and $\mathbb{D}$. Note that each variable set $v_h$ in either of the sets isolates one particular curve from the rest of the curves. This is intuitive since the best predictors maximize the dissimilarity between curves. 

First, we take a look at $v_3 \in \mathbb{G}$ (at 0600 hours) where the \emph{C}-curve is at its highest and narrowest (also when compared against the two other curves). At this hour, almost all children commuters are on their way to school. The narrowness of the \emph{C}-curve peak implies that the start time of schools are highly likely the same across the city-state and that they are more rigid than the adult working hours---the \emph{A}-curve within the same time frame is wider. In addition, at 0600 hours, the \emph{C}-curve is isolated from the intersecting \emph{A} and \emph{S}-curves. Almost similar dynamics is surmised for $v_5 \in \mathbb{D}$; however, the \emph{C}-curve peak has started to drop at lower travel demand levels. Second, we focus on $v_8 \in \mathbb{G}, \mathbb{D}$ at 1100 hours. At 1100 hours, both the students/children and the working adult population are in their schools/offices; that is, they are not traveling. This may explain why both \emph{A} and \emph{C}-curves at that hour are overlapping---isolating the \emph{S}-curve. In addition, notice that in Fig.~\ref{fig:ridestarttime}a, the \emph{S}-curve has no prominent peak unlike in the \emph{A}-curve (2 peaks) and \emph{C}-curve (1 peak). This is not surprising since most senior citizens do not follow ``regular'' adult working hours (although some may still do as depicted by the two shallow ``bumps'' around the same region where the \emph{A}-curve peaks). It can be said that, by and large, the elderlies do not have a ``universal'' schedule unlike the working adults, and that during ``working hours'' when the students and working adults are at work or in school, more elderlies are traveling. Finally, for $v_{11}$ and $v_{12}$ $\in \mathbb{G}, \mathbb{D}$, at 1400-1500 hours, the \emph{A}-curve is left at the lower levels of the travel demand and is isolated from the \emph{C} and \emph{S}-curves. This is a particularly interesting trend for the student/child population, which reveals that most students are only in schools for half a day and that they have varying end of school times. This is manifested in the \emph{C}-curve where there is no second peak observed as it starts to plateau at 1300 to 1800 hours. In addition, from 1400-1500 hours, the travel demand curve implies that most working adults are still in their offices. The insights presented here are summarized in Table~\ref{predictorsummary}.

\begin{table}[h]
\centering
\caption{\textbf{Top Predictor Variables}. We highlight the top predictor variables in terms of their variable importance values. It can be seen that for each predictor variable, a travel demand curve is isolated from the rest. Furthermore, all curves are, in one way or another, well-represented in the choice of feature variables.} 
\label{predictorsummary}
\begin{tabular}{llll}
\hline
\multicolumn{1}{c}{{\bf Predictor}} & \multicolumn{1}{c}{{\bf Hour of Day}} & \multicolumn{1}{c}{{\bf Isolated Curve}} & \multicolumn{1}{c}{{\bf Remarks}}                                                                                                                                                \\ \hline
\begin{tabular}[c]{@{}l@{}}$v_3$ \\ $v_5$ \end{tabular} & \begin{tabular}[c]{@{}l@{}}0600 hours   \\ 0800 hours  \end{tabular}                               & \emph{C}-Curve                                  & \begin{tabular}[c]{@{}l@{}}Students/children dominate travel demand\\  Travel demand highest and narrowest\end{tabular}                                                  \\ \hline
$v_8$ &1100 hours                                    & \emph{S}-Curve                                  & \begin{tabular}[c]{@{}l@{}}Working adults in offices \\ Children/students  in classes \\ Senior citizens traveling\end{tabular}                                     \\ \hline
\begin{tabular}[c]{@{}l@{}}$v_{11}$ \\ $v_{12}$ \end{tabular} & \begin{tabular}[c]{@{}l@{}}1400 hours \\ 1500 hours    \end{tabular}                                & \emph{A}-Curve                                  & \begin{tabular}[c]{@{}l@{}}Working adults in offices\\ Senior citizens traveling \\ Students/children traveling home\end{tabular} \\ \hline

\end{tabular}
\end{table}


\section{Conclusion}
A sufficient knowledge of the demographics of a commuting public is essential in formulating and implementing more targeted transportation policies---different schemes can affect different commuter types in several ways. In this work, using data taken from Singapore's automated fare collection (AFC) system, we showed that commuters exhibit varying travel patterns that can be used to categorize passengers into three general types: adult, child/student, senior citizen. We first established a method to construct distinct commuter matrices that we referred to as \emph{eigentravel} matrices that capture the characteristic travel routines of individuals by taking into account their times in the day of travel, durations of travel, and preferred modes of transport. We then performed a multivariate analysis (840 feature variables) on the eigentravel matrices using three supervised machine learning models: gradient boosting method (GBM), distributed random forest (DRF), and support vector mahine (SVM). GBM gave the best prediction accuracy of 76\%. Furthermore, implementing a variable importance analysis showed that features associated with weekday travels are better than those associated with weekends.

Many cities are already using AFC systems, and some metropolitan areas in the developing worlds are already transitioning into using such technology. However, not all AFC systems provide  passenger type information like what the dataset in this study provides. Nevertheless, with the approach presented, urban planners can now have a way to determine passenger types by looking at the ``natural tendencies'' of public transport commuters, thru eigentravel matrices, in a non-invasive manner. 

The technique demonstrated allows transport planners to formulate more targeted transportation policies and schemes. The framework is not only useful to urban and transport planners; the field of marketing research may also find this work relevant and beneficial. With adequate awareness of which passenger types dominate the travel demand at specific times of day, ad agencies (and survey firms) can create andt put up more focused advertisements, service announcements, and surveys---helping stakeholders to properly channel their resources. Finally, from the perspective of modeling and simulations, the categorization presented can be useful in generating synthetic populations for use as inputs to computational models (e.g. agent-based models) to accurately capture the revealed travel signatures for each commuter category.

\section*{Acknowledgement} 
\noindent
We would like to thank the Land Transport Authority of Singapore for the ticketing data used in this work, Nasri bin Othman for his assistance in preparing the datasets, and Hu Nan for his valuable feedback on the work. This research is supported by Singapore A$^*$SERC Complex Systems Programme research grant (1224504056). 



 \bibliographystyle{elsarticle-num} 
 \bibliography{myref}

\end{document}